\documentclass[prl,
                twocolumn,
                preprintnumbers,
                amsmath,
                amssymb,
                showpacs,
                superscriptaddress]{revtex4-1}
\usepackage{graphicx} 
\usepackage{dcolumn}  
\usepackage{bm}       
\usepackage{amsfonts}
\usepackage{dsfont}
\usepackage{mathtools}
\usepackage[colorlinks=true,linkcolor=blue,citecolor=red,urlcolor=blue]{hyperref}
\usepackage{color}
\newcommand{\eq}[1]{Eq.~(\ref{#1})}

\begin{document}

\title{Time delay in the recoiling valence-photoemission of Ar endohedrally confined in C$_{60}$}

\author{Gopal Dixit}
\email[]{dixit@mbi-berlin.de}
\affiliation{%
Center for Free-Electron Laser Science, DESY,
            Notkestrasse 85, 22607 Hamburg, Germany }
\affiliation{%
Max Born Institute, Max-Born-Strasse 2A, 12489 Berlin, Germany }
						
\author{Himadri S. Chakraborty}
\email[]{himadri@nwmissouri.edu}
\affiliation{%
Department of Natural Sciences, Center for Innovation and Entrepreneurship,
Northwest Missouri State University, Maryville, Missouri 64468, USA}
						
\author{Mohamed El-Amine Madjet}
\email[]{mmadjet@qf.org.qa}
\affiliation{%
Center for Free-Electron Laser Science, DESY,
            Notkestrasse 85, 22607 Hamburg, Germany }
\affiliation{%
Qatar Energy and Environment Research Institute (QEERI), Qatar Foundation, 
Doha, Qatar}

\date{\today}



\begin{abstract}
The effects of confinement and electron correlations on the relative time delay between the 3s and 3p photoemissions of
Ar confined endohedrally in C$_{60}$ are investigated using 
the time dependent local density approximation - a method that 
is also found to mostly agree with recent time delay measurements between the 3s and 3p subshells in atomic Ar. 
At energies in the neighborhood 
of 3p Cooper minimum, correlations with C$_{60}$ electrons are found 
to induce opposite temporal effects in the emission of Ar 3p 
hybridized symmetrically versus that of Ar 3p hybridized antisymmetrically with C$_{60}$. 
A recoil-type interaction model mediated
by the confinement is found to best describe the phenomenon.
\end{abstract}

\maketitle 
With the tremendous advancement in technology for generating attosecond ({\it as}) isolated pulses as well 
as attosecond pulse trains, it becomes possible to study fundamental phenomena of light-matter interaction 
with unprecedented precision on an {\it as} timescale \cite{hentschel, goulielmakis1, krausz}. 
In particular, the relative time delay between the 
photoelectrons from different subshells on {\it as} timescale, a subject of intense recent 
activities, is expected to probe important aspects of
electron correlations that predominantly influence the photoelectron.  
Pump-probe experiments have been performed to measure the relative 
delay in the photoemission processes, where extreme ultra-violet (XUV) pulses are used 
to remove an electron from a particular subshell and 
subsequently a weak infrared (IR) pulse accesses the temporal information of the emission event \cite{pazourek}. 

Streaking measurements 
were carried out to probe photoemission from the valence and the conduction band 
in single-crystalline magnesium \cite{neppl2012attosecond} 
and tungsten \cite{cavalieri2007attosecond}. 
A streaking technique was also employed to measure the relative delay of approximately 
21$\pm$5 {\it as} between the 2s and 2p subshells of atomic Ne at 106 eV photon energy \cite{schultze2010delay}. 
Despite several theoretical attempts \cite{mauritsson2005accessing, kheifets2010delay, 
moore2011time, ivanov2011accurate, nagele2012time, dahlstrom2012diagrammatic, kheifets2013time} 
to explain this measured delay in Ne, only about a half of the delay could be reproduced,
keeping the time delay in Ne photoemissions still an open problem. Recently, 
the relative delay between the 3s and 3p subshells in Ar is measured  at three 
photon energies by interferometric technique using attosecond 
pulses \cite{klunder2011probing, guenot2012photoemission}.
Theoretical methods (e.g. time-dependent nonperturbative method \cite{mauritsson2005accessing}, 
diagrammatic many-body perturbation theory 
\cite{dahlstrom2012diagrammatic}, Random phase approximation with exchange (RPAE) \cite{guenot2012photoemission, kheifets2013time}, 
and multi-configurational 
Hartree-Fock (MCHF) \cite{carette2013multiconfigurational}) have been employed to investigate this relative delay in Ar, 
although agreements between 
theory and experiment is rather inconclusive. 
A ubiquitous understanding in all these studies is the dominant
influence of electron correlations to determine the time behavior of outgoing electrons. Thus, it is fair to expect that
the process near a Cooper minimum or a resonance will be particularly nuanced.  

\begin{figure}[h!]
\includegraphics[width=7cm]{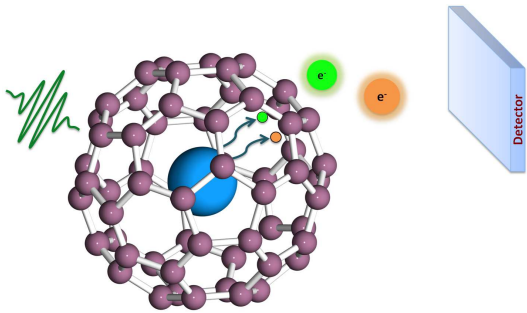}
\caption{(Color online). Schematic for probing the effects of correlations from the confinement 
on the relative time delay in the emission of an atom encaged endohedrally inside
C$_{60}$.} \label{fig1}
\end{figure}

It is therefore of  spontaneous interest to extend the study to test the effect of correlations 
on the temporal photoresponse of atoms in material 
confinements. A brilliant natural laboratory for such is an atom endohedrally captured in a fullerene shell;
see Fig.\,\ref{fig1} which envisions the process.  
There are two compelling reasons for this choice: (i) such materials are highly stable, 
have low-cost sustenance at room temperature and are enjoying a rapid improvement in their synthesis 
techniques \cite{popov2013} and (ii) effects of correlations of the central atom 
with the cage electrons have been predicted to spectacularly influence
the atomic valence photoionization \cite{madjet2007giant}. 
In this Letter, by considering Ar@C$_{60}$, we show that a 
confinement-induced correlation effect of C$_{60}$ at energies surrounding the 
Ar 3p Cooper minimum produces a faster and a slower emission of 
the Ar 3p electrons hybridized, respectively, in a symmetric and an antisymmetric mode with 
a near-degenerate C$_{60}$ orbital.

\begin{figure}[h!]
\includegraphics[width=8cm]{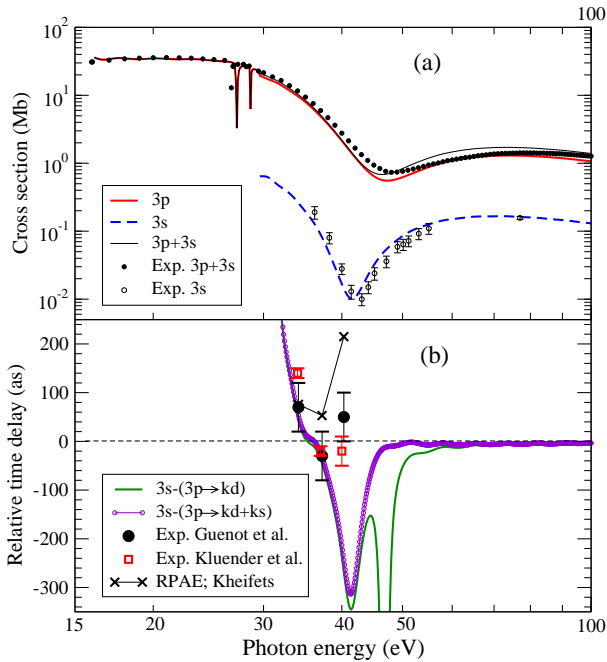}
\caption{(Color online).  Top: TDLDA 3p, 3s and total photoionization cross sections for atomic Ar 
are compared with experiments for 3s \cite{mobus1993measurements} and total \cite{samson2002precision}. 
For 3s the computed cross section is scaled to reproduce the measurement at the
Cooper minimum. Bottom: The relative TDLDA time delay between 3s and 3p of Ar and its comparison  
with measurements (solid black circles, Ref.\,\cite{guenot2012photoemission}; 
open red squares, Ref.\,\cite{klunder2011probing}). RPAE results \cite{kheifets2013time} at three experimental energies are also
cited.} \label{fig2}
\end{figure}

Time dependent local density approximation (TDLDA), with  Leeuwen and Baerends (LB) exchange-correlation functional  
to produce accurate asymptotic behavior \cite{van1994exchange} of ground and continuum electrons, is employed to
calculate the dynamical response of the system to the external electromagnetic field.    
To demonstrate the accuracy of 
the method for an isolated atom, the total photoionization cross section and the partial 3s and 3p cross sections 
of Ar are presented in Fig.\,\ref{fig2}a and compared with available 
experiments \cite{mobus1993measurements,samson2002precision}.
As seen, our TDLDA total and 3s cross sections are in excellent agreement with 
experimental results and the positions of the 3s and 3p Cooper minima at, respectively, 
42 and 48 eV are well reproduced. 
The dominance of 3p contribution over 3s in this energy range (Fig.\,\ref{fig2}a) 
also automatically implies the accuracy of our TDLDA 3p result. 

The absolute time delay in Ar pump-probe photoemission contains two contributions: 
one due to the absorption of XUV photon and the other due to the probe pulse. 
Owing to the weak probe pulse, the probe-assisted 
delay contributions can be estimated \cite{dahlstrom2012diagrammatic} as a function of the kinetic energy of electrons
from different Ar subshells. This allowed evaluation of the relative delay in recent measurements \cite{klunder2011probing, guenot2012photoemission}. This delay therefore connects to the energy derivative of the quantum phase of complex 
photoionization amplitude \cite{yakovlev2010attosecond} 
- the Wigner-Smith time delay \cite{wigner1955lower, smith1960lifetime, de2002time}.
Several methods \cite{ivanov2011accurate, nagele2012time, dahlstrom2012theory, ivanov2013extraction} 
have been utilized to extract the Wigner-Smith time delay directly from the measurements.

The photoionization amplitude from an initial bound state  ($n_{i}l_{i}$) to a 
final continuum state ($kl$) can be expressed as 
\begin{eqnarray}\label{eq5}
f(\hat{\bf{k}}) & = & (8 \pi)^{3/2} \sum_{\substack{l = l_{i} \pm1 \\ m = m_{i}}} (-i)^{l} e^{i \eta_{l}(\hat{\bf{k}})} Y_{lm}^{*}(\hat{\bf{k}})
\langle \phi_{kl} ||r+\delta V|| \phi_{n_{i}l_{i}} \rangle  \nonumber \\
&& \times \sqrt {(2l+1)(2l_{i}+1)}
\left (\begin{array}{ccc}
l & 1 & l_{i} \\
0 & 0 & 0
\end{array} \right)
\left (\begin{array}{ccc}
l & 1 & l_{i} \\
-m & 0 & m_{i}
\end{array} \right).
\end{eqnarray}
Here, $\delta V$ is the complex induced potential which embodies TDLDA many-body correlations. 
The phase $\eta_{l}$ includes contributions from both the
short range and Coulomb potentials, whereas the phase of the 
complex matrix element in \eq{eq5}
is the {\em correlation} phase. For Ar, the correlation near Cooper minima primarily arises from the coupling 
of 3p with 3s channels. 
The total phase is the sum of these three contributions. 
The time delay profile is computed by differentiating the TDLDA total phase in energy. 

Our TDLDA relative Wigner-Smith delay between Ar 3s and 3p, $\tau_{\textrm{3s}}-\tau_{\textrm{3p}}$, 
is compared with the experimental data 
of Gu{\'e}not {\it et al.}\,\cite{guenot2012photoemission} and of Kl{\"u}nder {\it et al.}\,\cite{klunder2011probing} 
in Fig.\,\ref{fig2}b. 
As seen, the relative delay is strongly energy dependent. 
Note that the TDLDA results are in excellent agreement with both sets of 
experimental results at  34.1 and 37.2 eV.
The third measurement at 40.3 eV, which is in the vicinity of the 3s Cooper minimum, is negative in 
Ref.\,\cite{klunder2011probing} in contrast to its positive
value in Ref.\,\cite{guenot2012photoemission}. Note that our result captures the correct sign as 
in Kl{\"u}nder {\it et al.} at 40.3 eV.
In general, 3p$\rightarrow$kd photochannel is dominant over 3p$\rightarrow$ks at most energies. 
Close to the 3p Cooper minimum, however, 3p$\rightarrow$kd 
begins to rapidly decrease to its minimum value, enabling 3p$\rightarrow$ks to 
significantly contribute to the {\em net} 3p delay. The s- and d-wave emissions have different angular distributions but 
their Wigner delays are independent of emission directions. Thus, assuming that {\em all} 3p photoelectrons are detected (integration over solid angle),
the net 3p delay must be a statistical combination, that is, the sum of the delays weighted by the channel's
individual cross section branching ratios.
As illustrated in Fig.\,\ref{fig2}b, upon including 3p$\rightarrow$ks along with 3p$\rightarrow$kd (purple curve) 
this way, the shape of the TDLDA delay strikingly alters near 3p Cooper minimum.  We stress that the delay near a 
Cooper minimum needs to be addressed with great care which can reveal new physics, 
as shown below for an endohedrally confined Ar atom. 

We also include recent RPAE results \cite{kheifets2013time} for three experimental energies in Fig.\,\ref{fig2}b. 
As seen, RPAE and experiments match only at 34.1 eV. The superior performance of TDLDA in explaining the measurements is thus evident.
While both TDLDA and RPA are many-body linear response theories, they have significant differences in the details, particularly,
in treating electron correlations \cite{onida2002electronic}. 
Variants of the Kohn-Sham LDA+LB scheme were successfully utilized to describe attosecond strong-field phenomena~\cite{petretti2010alignment, heslar2011high, farrell2011strong, toffoil2012, hellgren2013}, underscoring
the reliability of many-body correlations that TDLDA characteristically offers.

This success of TDLDA method for free Ar encouraged us to use the approach 
to investigate the delay in an Ar atom endohedrally sequestered
in C$_{60}$. 
The jellium model is employed for computing the relative delay \cite{madjet2010}. This model 
enjoyed earlier successes in codiscovering with experimentalists a high-energy plasmon 
resonance \cite{scully2005}, interpreting the energy-dependent oscillations 
in C$_{60}$ valence photo-intensities \cite{rudel2002}, and predicting giant enhancements in the 
confined atom's photoresponse from the coupling with C$_{60}$ plasmons \cite{madjet2007giant}. 
Significant ground state hybridization of Ar 3p
is found to occur with the C$_{60}$ 3p orbital, resulting in  
3p[Ar+C$_{60}$] and  3p[Ar-C$_{60}$] from,
respectively, the symmetric and antisymmetric wavefunction mixing. These are spherical analogs of bonding and 
antibonding states in molecules or dimers. Such atom-fullerene hybridization was predicted 
before \cite{chakraborty2009} and detected in the photoemission experiment on 
multilayers of Ar@C$_{60}$ \cite{morscher2010strong}. In fact, the hybridization gap of 
1.5 eV between 3p[Ar+C$_{60}$] and 3p[Ar-C$_{60}$] in our calculation is in good agreement 
with the measured value of 1.6$\pm$0.2 eV \cite{morscher2010strong}.    

\begin{figure}[h!]
\vspace{1.0 cm}
\includegraphics[width=7cm]{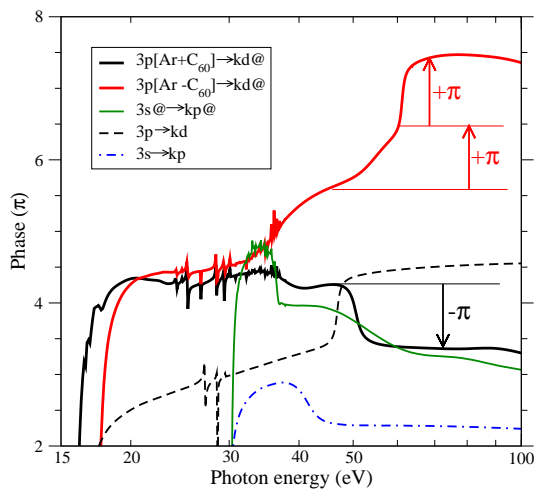}
\caption{(Color online). TDLDA quantum phases for ionization via d-waves from bonding 3p[Ar+C$_{60}$] and 
antibonding 3p[Ar-C$_{60}$] levels and via p-wave from Ar 3s@ are compared with their counterparts in free Ar.} 
\label{fig3}
\end{figure}

The TDLDA Wigner-Smith phases for relevant ionization channels for confined and free Ar are presented in Fig.\,\ref{fig3}. 
We use the symbol ``@'' to denote states belonging to the confined Ar. The narrow resonance spikes below 40 eV 
are due to single electron Rydberg-type excitations in C$_{60}$. This energy zone also includes 
the C$_{60}$ plasmon resonances, although their effects are suppressed by the Coulomb phase that dominates
the extended region above ionization thresholds. 
We note that the Ar 3s Cooper minimum shifts slightly lower in energy to 36.5 eV from the confinement, but the confinement moves the two 3p minima, 
each in the bonding and antibonding channels, somewhat higher in energy. 
What is rather dramatic in Fig.\,\ref{fig3} is 
that the quantum phase corresponding to  3p[Ar+C$_{60}$]$\rightarrow$kd@ (thick solid black) makes a downward $\pi$ phase shift, whereas 
the phase associated with 3p[Ar-C$_{60}$]$\rightarrow$kd@ (thick solid red) suffers a upward 2$\pi$ phase shift at their respective Cooper minimum. 
Further note that both these contributions together yield a {\em net} phase that shifts up by $\pi$ as in the case of 
free-Ar 3p$\rightarrow$kd channel (dashed black curve in Fig.\,\ref{fig3}) at its Cooper minimum.

This contrasting phase behavior between hybrid 3p emissions is likely the effect of symmetric and antisymmetric wavefunction shapes
on the matrix elements through dynamical correlations.
Using the well-known Fano scheme of perturbative interchannel coupling \cite{fano1961} the lead contribution 
to the matrix element $\langle\delta V\rangle$ (\eq{eq5}) is \cite{javani2012}
\begin{equation}\label{matelm}
\langle\delta V\rangle_{\alpha} (E) = \displaystyle\sum_{\beta}\int dE' \frac{\langle\Psi_{\beta}(E')|\frac{1}{|{\bf r}_{\alpha}-{\bf r}_{\beta}|}
|\Psi_{\alpha}(E)\rangle}{E-E'} \langle z\rangle_{\beta} (E'),
\end{equation}
where $\alpha$ denotes each of the 3p[Ar$\pm$ C$_{60}$]$\rightarrow$kd@ channels. $\Psi$ are {\em channel}-wavefunctions that involve both bound (hole) and continuum (photoelectron) states, 
and $\langle z \rangle_{\beta}$ is the single channel matrix element of each perturbing channel $\beta$. 
Thus, the summation over channels incorporates bound states as the hole states.
Two points can be noted: First,  $\langle\delta V\rangle$ 
dominates near the Cooper minimum of a channel $\alpha$,
since the ``unperturbed" $\langle z \rangle_{\alpha}$ is already small at these energies;
second, $\langle\delta V\rangle$ 
depends on the coupling matrix element in the numerator of \eq{matelm} that involves overlaps between the bound state of a $\alpha$ 
channel with that in a perturbing $\beta$ channel. These overlaps are critical, since 3p[Ar+C$_{60}$] wavefunction has a structure
completely opposite to that of 3p[Ar-C$_{60}$] over the C$_{60}$ shell region where each of them strongly overlaps with
a host of C$_{60}$ wavefunctions to build correlations.
These opposing modes of overlap from one hybrid to another flip the phase modification direction 
between two hybrid 3p emissions around a respective Cooper minimum, as seen in Fig.\,\ref{fig3}.
 
\begin{figure}[ht]
\includegraphics[width=8cm]{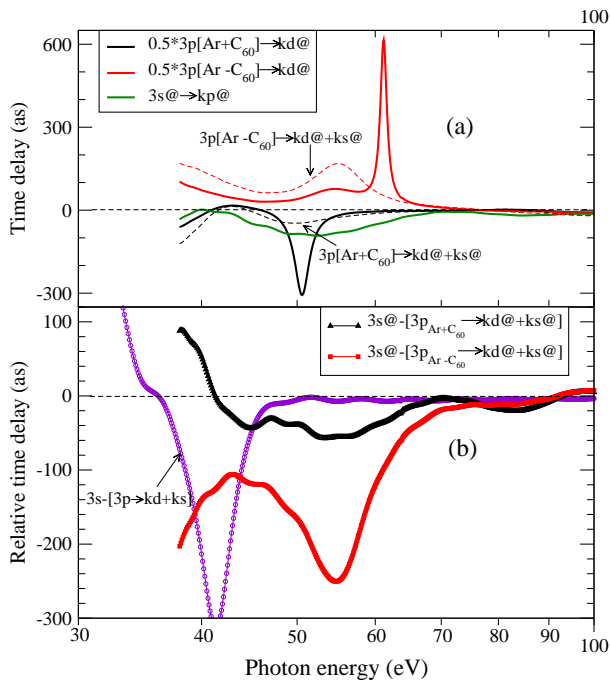}
\caption{(Color online). Top: Absolute time delay for ionizations in 3p[Ar$\pm$C$_{60}$]$\rightarrow$kd@ 
and 3s@$\rightarrow$kp@ channels. For the two hybrid channels, results modified by incorporating s-wave 
delays are also shown.  
Bottom: Relative delays $\tau_{\textrm{3s@}}-\tau_{\textrm{3p [Ar} \pm \textrm{C}_{60}]}$, 
including the s-wave contributions; $\tau_{\textrm{3s}}-\tau_{\textrm{3p}}$ of free Ar is also shown for comparison.
} 
\label{fig4}
\end{figure}

Depending on the upward (downward) shift in the quantum phase, the resulting photoelectron exhibits positive (negative) time delay 
and hence emerges slower(faster) from the ionization 
region. This is evident in Fig.\,\ref{fig4}a, which features various absolute delays: 
Channels 3p[Ar+C$_{60}$]$\rightarrow$ kd@ and
3p[Ar-C$_{60}$]$\rightarrow$ kd@ exhibit, respectively, a fast and a slow emission over relatively narrow ranges 
about their Cooper minima.
Note that the peak delay of the antibonding electron is approximately double to the peak 
advancement (negative delay)
of the bonding electron. 
The delay profile  becomes softer and broader in energy by including the contribution from s-wave, but the general trend of a rapid and a 
slow ejection, respectively, in the bonding and antibonding channels survives.

The conservation of the quantum phase, i.e., the net phase shift of $\pi$ in the upward direction (as in the free Ar) for
3p in Ar@C$_{60}$, can be understood in the language of a collision type interaction between two hybrid 3p electrons.
The phase behaves like the linear momentum in a two-body collision which is a conserved quantity. Its energy derivative, i.e., 
the time delay, can be thought to be commensurate with the collision force, the time derivative of the momentum, 
since time and energy are conjugate variables. This implies, that if one hybrid electron goes 
through an advanced emission, the other hybrid must delay or {\em time-recoil} appropriately to keep the net delay 
roughly close to the delay of free Ar. Of course, here the process is underpinned by the orbital mixing. 
Therefore, the phenomenon can be pictured as the photo-liberation of two recoiling electrons in the temporal domain 
from the atom-fullerene hybridization. Hence, it is also likely to exist in the ionization of molecules, nanodimers, and
fullerene onions that support hybrid electrons.

The time delays in the photoionization of 3p hybrids (with s-wave contribution included) relative to 3s@, 
$\tau_{\textrm{3s@}}-\tau_{\textrm{3p [Ar} \pm \textrm{C}_{60}]}$, are presented in Fig.\,\ref{fig4}b. 
One notes in Fig.\,\ref{fig4}a that 
3s@$\rightarrow$kp@ produces an absolute delay profile, 
which is negative for most energies and, on an average, comparable to the absolute delay in 
3p[Ar-C$_{60}$]$\rightarrow$kd@+ks@. 
Consequently, their (fast) emergence at about similar speeds keeps their relative delay close to 
$\tau = 0$, but with a bias toward negative values. 
On the other hand, for the 3p[Ar+C$_{60}$]$\rightarrow$kd@+ks@ channel the relative delay remains mostly strongly negative. 
However, the rich structures in the delay profiles emphasize that the Cooper 
minimum regions are particularly attractive for time delay studies.

The 3p bonding-antibonding gap of 1.5 eV requires the energy of the probe
pulse to be smaller than this gap. Otherwise, the sideband of one level will begin to overlap with the harmonics of the other. 
Also, by varying the polarization angle between XUV and IR pulses one can potentially probe both independent contributions, i.e., the relative delay between 3s orbital and 3p bonding/anti-bonding orbital, i.e., by extending the standard RABBIT method ~\cite{veniard1995}, where 
the polarization of XUV pulse is the same as the IR pulse. 
Therefore, techniques based on interferometry, such as RABBIT 
and PROOF \cite{chini2010}, have potentials
to probe the relative delay between 3p bonding/antibonding and 3s electrons. One may also perform the streaking experiments using IR as well as THz pulses for accessing the delay. 
We suggest that future experiments be performed on the time delay in Ar and Ar@C$_{60}$ over broader photon energy 
ranges including the 3p Cooper minimum to unravel new physics from confinement and correlations.

In conclusion, our TDLDA relative Wigner-Smith time delay between 3s and 3p subshells in free Ar are
in excellent agreement with the measured delay except near the 3s Cooper minimum, where, however, the TDLDA
is consistent with the sign of one set of measurements.
In the case of confined Ar, due to the electron correlation, the delays of the 3p bonding and 3p antibonding emissions
are governed by a recoil-type emission in the time-domain mediated by the host C$_{60}$. It is found that the
emission from the 3s@ level is slightly faster than the emission from the 3p bonding level but 
is substantially faster, by 100 {\it as} and above, than 
the emission from the 3p antibonding level.
We further demonstrate that the delay of Ar 3p electron, free or confined, leads to significant modifications in
the vicinity of the Cooper minimum by including the s-wave photochannel.

\begin{acknowledgments}
G. D. acknowledges Misha Ivanov, Tim Laarmann and Oliver M\"{u}cke for useful discussions. 
The research is supported by the NSF, USA.
\end{acknowledgments}


\begin{thebibliography}{10}

\bibitem{hentschel}
M.~Hentschel, R.~Kienberger, C.~Spielmann, G.~A. Reider, N.~Milosevic,
  T.~Brabec, P.~Corkum, U.~Heinzmann, M.~Drescher, and F.~Krausz,
\newblock Nature {\bf 414}, 509 (2001).

\bibitem{goulielmakis1}
E.~Goulielmakis, M.~Schultze, M.~Hofstetter, V.~S. Yakovlev, J.~Gagnon,
  M.~Uiberacker, A.~L. Aquila, E.~M. Gullikson, D.~T. Attwood, R.~Kienberger,
  F.~Krausz, and U.~Kleineberg,
\newblock Science {\bf 320}, 1614 (2008).

\bibitem{krausz}
F.~Krausz and M.~Ivanov,
\newblock Rev. Mod. Phys. {\bf 81}, 163 (2009).

\bibitem{pazourek}
R.~ Pazourek, S.~ Nagele, and J.~ Burgd\"{o}rfer,
\newblock Faraday Discuss. {\bf 163}, 353 (2013).


\bibitem{neppl2012attosecond}
S.~Neppl, R.~Ernstorfer, E.~M. Bothschafter, A.~L. Cavalieri, D.~Menzel, J.~V.
  Barth, F.~Krausz, R.~Kienberger, and P.~Feulner,
\newblock Physical Review Letters {\bf 109}, 87401 (2012).

\bibitem{cavalieri2007attosecond}
A.~L. Cavalieri, N.~Müller, T.~Uphues, V.~S. Yakovlev, A.~Baltuska, B.~Horvath,
  B.~Schmidt, L.~Blümel, R.~Holzwarth, S.~Hendel, M.~Drescher, U.~Kleineberg,
  P.~M. Echenique, R.~Kienberger, F.~Krausz, and U.~Heinzmann,
\newblock Nature {\bf 449}, 1029 (2007).

\bibitem{schultze2010delay}
M.~Schultze, M.~Fieß, N.~Karpowicz, J.~Gagnon, M.~Korbman, M.~Hofstetter,
  S.~Neppl, A.~L. Cavalieri, Y.~Komninos, T.~Mercouris, C.~A. Nicolaides,
  R.~Pazourek, S.~Nagele, J.~Feist, J.~Burgdörfer, A.~M. Azzeer, R.~Ernstorfer,
  R.~Kienberger, U.~Kleineberg, E.~Goulielmakis, K.~F., and Y.~V. S.,
\newblock Science {\bf 328}, 1658 (2010).

\bibitem{mauritsson2005accessing}
J.~Mauritsson, M.~B. Gaarde, and K.~J. Schafer,
\newblock Physical Review A {\bf 72}, 013401 (2005).

\bibitem{kheifets2010delay}
A.~S. Kheifets and I.~A. Ivanov,
\newblock Physical Review Letters {\bf 105}, 233002 (2010).

\bibitem{moore2011time}
L.~R. Moore, M.~A. Lysaght, J.~S. Parker, H.~W. van~der Hart, and K.~T. Taylor,
\newblock Physical Review A {\bf 84}, 061404 (2011).

\bibitem{ivanov2011accurate}
M.~Ivanov and O.~Smirnova,
\newblock Physical Review Letters {\bf 107}, 213605 (2011).

\bibitem{nagele2012time}
S.~Nagele, R.~Pazourek, J.~Feist, and J.~Burgd{\"o}rfer,
\newblock Physical Review A {\bf 85}, 033401 (2012).

\bibitem{dahlstrom2012diagrammatic}
J.~M. Dahlstr{\"o}m, T.~Carette, and E.~Lindroth,
\newblock Physical Review A {\bf 86}, 061402 (2012).

\bibitem{kheifets2013time}
A.~S. Kheifets,
\newblock Physical Review A {\bf 87}, 063404 (2013).

\bibitem{klunder2011probing}
K.~Kl{\"u}nder, J.~M. Dahlström, M.~Gisselbrecht, T.~Fordell, M.~Swoboda,
  D.~Guénot, P.~Johnsson, J.~Caillat, J.~Mauritsson, A.~Maquet, R.~Taïeb, and
  A.~L'Huillier,
\newblock Physical Review Letters {\bf 106}, 143002 (2011).

\bibitem{guenot2012photoemission}
D.~Gu{\'e}not, K.~Kl{\"u}nder, C.~L. Arnold, D.~Kroon, J.~M. Dahlstr{\"o}m,
  M.~Miranda, T.~Fordell, M.~Gisselbrecht, P.~Johnsson, J.~Mauritsson,
  E.~Lindroth, A.~Maquet, R.~Ta{\"\i}eb, A.~L`Huillier, and A.~S. Kheifets,
\newblock Physical Review A {\bf 85}, 053424 (2012).

\bibitem{carette2013multiconfigurational}
T.~Carette, J.~M. Dahlstr{\"o}m, L.~Argenti, and E.~Lindroth,
\newblock Physical Review A {\bf 87}, 023420 (2013).

\bibitem{popov2013}
A.~A. Popov, S.~Yang, and L.~Dunsch,
\newblock Chem. Rev. {\bf 113}, 5989--6113 
  (2013).

\bibitem{madjet2007giant}
M.~E. Madjet, H.~S. Chakraborty, and S.~T. Manson,
\newblock Physical Review Letters {\bf 99}, 243003 (2007).

\bibitem{van1994exchange}
R.~Van~Leeuwen and E.~J. Baerends,
\newblock Physical Review A {\bf 49}, 2421 (1994).

\bibitem{mobus1993measurements}
B.~M{\"o}bus, B.~Magel, K.~H. Schartner, B.~Langer, U.~Becker, M.~Wildberger,
  and H.~Schmoranzer,
\newblock Physical Review A {\bf 47}, 3888 (1993).

\bibitem{samson2002precision}
J.~A.~R. Samson and W.~C. Stolte,
\newblock Journal of electron spectroscopy and related phenomena {\bf 123}, 265
  (2002).

\bibitem{yakovlev2010attosecond}
V.~S. Yakovlev, J.~Gagnon, N.~Karpowicz, and F.~Krausz,
\newblock Physical Review Letters {\bf 105}, 73001 (2010).

\bibitem{wigner1955lower}
E.~P. Wigner,
\newblock Physical Review {\bf 98}, 145 (1955).

\bibitem{smith1960lifetime}
F.~T. Smith,
\newblock Physical Review {\bf 118}, 349 (1960).

\bibitem{de2002time}
C.~A.~A. de~Carvalho and H.~M. Nussenzveig,
\newblock Physics Reports {\bf 364}, 83 (2002).

\bibitem{dahlstrom2012theory}
J.~M. Dahlstr{\"o}m, D.~Gu{\'e}not, K.~Kl{\"u}nder, M.~Gisselbrecht,
  J.~Mauritsson, A.~L`Huillier, A.~Maquet, and R.~Ta{\"\i}eb,
\newblock Chemical Physics  (2012).

\bibitem{ivanov2013extraction}
I.~A. Ivanov and A.~S. Kheifets,
\newblock Physical Review A {\bf 87}, 063419 (2013).

\bibitem{onida2002electronic}
G.~Onida, L.~Reining, and A.~Rubio,
\newblock Reviews of Modern Physics {\bf 74}, 601 (2002).

\bibitem{petretti2010alignment}
S.~Petretti, Y.~V. Vanne, A.~Saenz, A.~Castro, and P.~Decleva,
\newblock Physical Review Letters {\bf 104}, 223001 (2010).

\bibitem{heslar2011high}
J.~Heslar, D.~Telnov, and S.~I. Chu,
\newblock Physical Review A {\bf 83}, 043414 (2011).

\bibitem{farrell2011strong}
J.~P. Farrell, S.~Petretti, J.~F{\"o}rster, B.~K. McFarland, L.~S. Spector,
  Y.~V. Vanne, P.~Decleva, P.~H. Bucksbaum, A.~Saenz, and M.~G{\"u}hr,
\newblock Physical Review Letters {\bf 107}, 083001 (2011).

\bibitem{toffoil2012}
D.~Toffoil, and P.~Decleva,
\newblock J. Chem. Phys. {\bf 137}, 134103 (2012).

\bibitem{hellgren2013}
M.~Hellgren, E.~Rasanen, and E.~K. U. Gross,
\newblock arXiv:1309.0786 (2013).

\bibitem{madjet2010}
M.~E. Madjet, T.~Renger, D.~E. Hopper, M.~A. McCune, H.~S. Chakraborty, J.~M.
  Rost, and S.~T. Manson,
\newblock Physical Review A {\bf 81}, 013202 (2010).

\bibitem{scully2005}
S.~W.~J. Scully, E.~D. Emmons, M.~F. Gharaibeh, R.~A. Phaneuf, A.~L.~D.
  Kilcoyne, A.~S. Schlachter, S.~Schippers, A.~M{\"u}ller, H.~S. Chakraborty,
  M.~E. Madjet, and J.~M. Rost,
\newblock Physical Review Letters {\bf 94}, 065503 (2005).

\bibitem{rudel2002}
A.~R{\"u}del, R.~Hentges, U.~Becker, H.~S. Chakraborty, M.~E. Madjet, and J.~M.
  Rost,
\newblock Physical Review Letters {\bf 89}, 125503 (2002).

\bibitem{chakraborty2009}
H.~S. Chakraborty, M.~E. Madjet, T.~Renger, J.~M. Rost, and S.~T. Manson,
\newblock Physical Review A {\bf 79}, 061201 (2009).

\bibitem{morscher2010strong}
M.~Morscher, A.~P. Seitsonen, S.~Ito, H.~Takagi, N.~Dragoe, and T.~Greber,
\newblock Physical Review A {\bf 82}, 051201 (2010).

\bibitem{fano1961}
U.~Fano,
\newblock Physical Review {\bf 124}, 1866 (1961).

\bibitem{javani2012}
M.~H. Javani, M.~R. McCreary, A.~B. Patel, M.~E. Madjet, H.~S. Chakraborty, and
  S.~T. Manson,
\newblock The European Physical Journal D {\bf 66}, 189 (2012).

\bibitem{veniard1995}
V.~Veniard, R.~Taieb, and A.~Maquet, 
\newblock  Physical Review Letters {\bf 74}, 4161 (1995).

\bibitem{chini2010}
M.~ Chini, S.~Gilbertson, S.~D.~Khan, and Z. Chang,  
\newblock Opt. Express {\bf 18}, 13006 (2010).

\end{thebibliography}

\end{document}